# REACTIVITY MONITORING USING THE AREA METHOD FOR THE SUBCRITICAL VENUS-F CORE WITHIN THE FRAMEWORK OF THE FREYA PROJECT

N. Marie[1], G. Lehaut[1], J.-L. Lecouey[1], A. Billebaud[2], S. Chabod[2], X. Doligez[3], F.-R. Lecolley[1], A. Kochetkov[4], W. Uyttenhove[4], G. Vittiglio[4], J. Wagemans[4], F. Mellier[5], G. Ban[1], H.-E. Thyébault[2], D. Villamarin[6]

[1] Laboratoire de Physique Corpusculaire de Caen, ENSICAEN/Univ. de Caen/CNRS-IN2P3, France
[2] Laboratoire de Physique Subatomique et de Cosmologie, CNRS-IN2P3/UJF/INPG, France
[3] Institut de Physique Nucléaire d'Orsay, CNRS-IN2P3/Univ. Paris Sud, France
[4] StudieCentrumvoorKernenergie-Centre d'Etude de l'Energie Nucléaire, Belgium
[5] Commissariat à l'Energie Atomique et aux Energies Alternatives, DEN/DER/SPEX, France
[6] Centro de InvestigationesEnergeticasMedioAmbientales y Tecnologicas, Spain

*Onbehalf of the FREYA collaboration*

## Abstract

Accelerator-Driven Systems (ADS) could be employed to incinerate minor actinides and so partly contribute to answer the problem of nuclear waste management. An ADS consists of the coupling of a subcritical fast reactor to a particle accelerator via a heavy material spallation target. The on-line reactivity monitoring of such an ADS is a serious issue regarding its safety.

In order to study the methodology of this monitoring, zero-power experimentswere undertaken at the GUINEVERE facility within the framework of the FP6-IP-EUROTRANS programme. Such experiments have been under completion within the FREYA FP7 project. The GUINEVERE facility is hosted at the SCK-CEN site in Mol (Belgium). It couples the VENUS-F subcritical fast core with the GENEPI-3C accelerator. The latter delivers a beam of deuterons, which are converted into 14-MeV neutrons via fusion reactions on a tritiated target.

This paper presents one of the investigated methods for ADS on-line reactivity monitoring which has to be validated in the program of the FREYA project. It describes the results obtained when Pulsed Neutron Source experiments are analysed using the so called Area Method, in order to estimate the reactivity of a few sub-critical configurations of the VENUS-F reactor, around $k_{eff}= 0.96$.

First the GUINEVERE facility is described. Then, following general considerations on the Area method, the results of its application to the neutron population time decrease spectra measured after a pulse by several fission chambers spread out over the whole reactor are discussed. Finally the reactivity values extracted are compared to the static reactivity values obtained using the Modified Source Multiplication (MSM) method.

**Introduction**

Accelerator-Driven Systems (ADS) could be employed to incinerate minor actinides and so partly contribute to answer the problem of nuclear waste management. An ADS consists of the coupling of a subcritical fast reactor to an accelerator whose light ion beam hits a heavy material spallation target immersed inside its lead alloy cooled core, so as to provide the extra external neutrons needed to sustain the power delivered by the reactor core. The on-line reactivity monitoring of such an ADS is a serious issue regarding its safety.

In order to study the methodology of this monitoring, zero-power experiments were initiated within the framework of the MUSE programme (FP5) [1] and further developed within the GUINEVERE (Generation of Uninterrupted Intense NEutron pulses at the lead VEnusREactor) project[2] of the FP6-IP-EUROTRANS programme [3]. Such experiments have been under completion within the FP7 FREYA (Fast Reactor Experiments for hYbrid Applications) project [4].

The GUINEVERE facility is hosted at the SCK-CEN site in Mol (Belgium). It is the result of the coupling of the VENUS-F subcritical fast core, composed of enriched uranium and solid lead, with the GENEPI-3C accelerator delivering a deuteron beam which impinges on a Tritium target installed at the reactor core center. The 14-MeV neutrons produced by the T(d,n) fusion reactions provide the external neutron source. The latter can be operated in a pulsed mode or in a continuous mode with periodic short beam interruptions, referred to as "beam trips".

This paper presents one of the investigated methods for ADS on-line reactivity monitoring which has to be validated in the program of the FREYA project. It describes the results obtained when Pulsed Neutron Source experiments are analysed using the so called Area Method [5], in order toestimate the reactivity of a fewsub-critical configurations of the VENUS-F reactor, around $k_{eff}$ = 0.96 (a typical configuration among the ones of interest for ADS studies). This technique could be exploited during core loading and start-up phases of an ADS.

First the GUINEVERE facility is described. Then, following general considerations on the Area method, the results of its application to the neutron population time decrease spectra measured after a pulse by several fission chambers spread out over the whole reactorare discussed. Finally the reactivity values extracted are compared to the static reactivity values obtained using the Modified Source Multiplication (MSM) method.

**The Guinevere facility**

Initially, the VENUS facility, located at SCK-CEN,Mol (Belgium), was a critical water-moderated thermal reactor. It was modified to become a fast reactor with highly enriched metal uranium and lead, further on referred to as VENUS-F (Fig.1). It can be coupled to an accelerator, GENEPI-3C, which delivers a deuteron beam (at about 220 keV energy), either in a continuous mode (with and without beam interruptions) or in a pulsed mode. The beam impinging on a copper target with a titanium-tritium (TiT) deposit, provides 14-MeV neutrons via T(d,n)$^4$He reactions, right at the center of the VENUS-F core.

*The GENEPI-3C accelerator*

On the contrary to an industrial ADS, the GUINEVERE neutron source is not provided by high energy spallation reactions but by T(d,n)$^4$He fusion reactions by means of the accelerator GENEPI-3C (GEnérateur de NEutronsPulsé et Intense) [2]. Built by a collaboration of CNRS-IN2P3 laboratories and first assembled at the Laboratoire de Physique Subatomiqueet de Cosmologie (Grenoble, France), it accelerates deuteron ions to the energy of 220 keV and guides them onto a tritiated target. In the GUINEVERE facility, the target is located at the core mid-plane of the VENUS-F reactor. This source provides a quasi-isotropic field of about 14 MeV neutrons.

This accelerator was designed for the GUINEVERE program and has dedicated specifications. In pulsed mode the GENEPI-3C accelerator provides one-microsecond pulses of around 20 mA peak current. The neutron source intensity in this mode is around $1-2 \times 10^6$ neutrons/pulse.

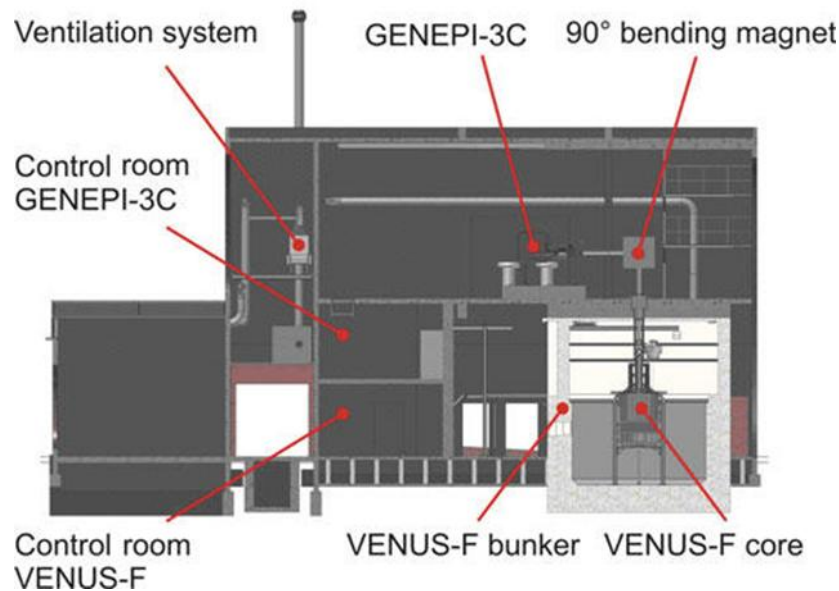

**Figure 1.** Sketch of the GUINEVERE facility.

*The VENUS-F reactor*

The VENUS-F fast zero power reactor takes place in a cylindrical vessel of approximately 80 cm in radius and 140 cm in height. A 12x12 grid surrounded by a 30 mm stainless steel casing can receive up to 144 elements of 8x8 $cm^2$ in section, which currently can be fuel assemblies, lead assemblies or specific elements for accommodating detectors or absorbent rods. The remaining room in the vessel is filled with semi-circular lead plates, which act as an outer radial neutron reflector. In addition, the core is reflected by top and bottom 40 cm-thick lead reflectors. Each fuel assembly (FA) consists of a 5x5 pattern filled with 9 fuel rodlets and 16 lead bars surrounded by lead plates. The fuel is 30 wt. % enriched metallic uranium provided by the CEA.

Various configurations of the reactor in terms of reactivity can be studied thanks to the modular shape of the core. The main configurations of interest herein are all derived from the so-called SC1 subcritical configuration shown in Fig. 2. 93 FAs (dark gray) are arranged in a way to create a pseudo-cylindrical core. Among them, six are actually safety rods (SR) made of boron-carbide with fuel followers with the absorbent part retracted from the core in normal operation. At the core periphery two boron-carbide control rods (CR, light gray) are used to adjust the reactivity. They can be moved from 0 mm (fully inserted inside the core) to 600 mm (fully retracted). For the SC1 configuration, both CRs are at 479.3 mm. The so-called PEAR (Pellet Absorber Rod) rod (light gray) is used for rod drop experiments. Its reactivity worth is very small (-136±5 pcm [6]) and it can be dropped almost instantaneously (in less than 0.5 second). It is fully inserted when the reactor is in the SC1 configuration. The remaining slots in the 12x12 grid are filled with pure lead assemblies (very light gray).

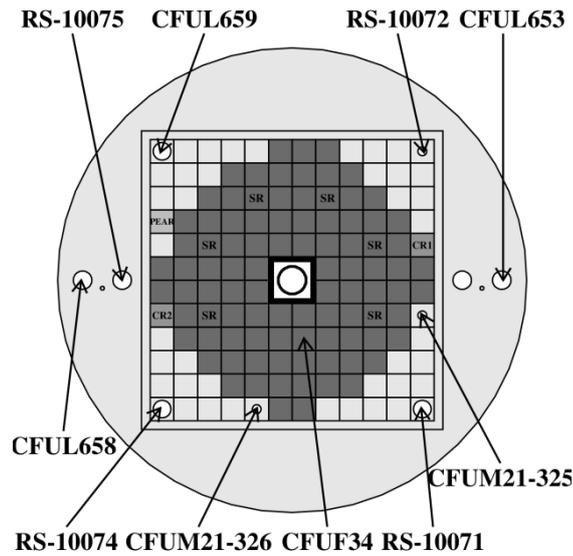

**Figure 2.** Cross view of the SC1 configuration.

SC1 was the first configuration to be studied in the foreseen dynamical reactivity measurement experiments. By moving the two control rods around their initial position, that is 479.3 mm, the other subcritical reactor configurations, also studied in this paper, were obtained. There are named SC1/CR=0 mm and SC1/CR=600mm.

In order to study the evolution of the neutron population in the reactor after injection of neutron pulses at the core center, ten fission chambers (FC) with $^{235}$U deposit were installed in the reactor. Table 1 gathers the detector names and their compositions as well as their deposit masses. For practical issues (presence of the guiding structure of the vertical beam line and, safety and control rod mechanisms) but also owing to experimental requirements (interest for a homogeneous fissile zone without local perturbation) all the detectors except one have been positioned in the reflector as shown in Fig. 2.

**Table 1.** Detectors used for the PNS experiments.

| Detector | Deposit | ~Mass (mg) |
|---|---|---|
| CFUL659 | $^{235}$U ($\approx$92%) | 1000 |
| CFUL658 | $^{235}$U ($\approx$92%) | 1000 |
| CFUL653 | $^{235}$U ($\approx$92%) | 1000 |
| RS-10071 | $^{235}$U ($\approx$90%) | 100 |
| RS-10072 | $^{235}$U ($\approx$90%) | 100 |
| RS-10074 | $^{235}$U ($\approx$90%) | 100 |
| RS-10075 | $^{235}$U ($\approx$90%) | 100 |
| CFUF34 | $^{235}$U ($\approx$100%) | 1 |
| CFUM21-325 | $^{235}$U ($\approx$90%) | 10 |
| CFUM21-326 | $^{235}$U ($\approx$90%) | 10 |

In order to test the performances of the Area method for reactivity monitoring, the reactivity of each subcritical configuration was first determined by other experiments using the MSM (Modified Source Multiplication) method [6]. It is a well-established static reactivity measurement technique, which has been extensively and successfully used to determine large subcriticality levels (up to several dollars). The unknown reactivity is determined by comparing detector count rates driven by an external neutron source in the configuration of interest with those obtained in another subcritical configuration whose reactivity is known [7]. Indeed, SC1 was first obtained from a critical configuration CR0 by removing the four central fuel assemblies (which allows inserting the accelerator beam tube) and by dropping the PEAR rod. A slightly subcritical configuration of known reactivity was created by simply dropping the PEAR rod in the reactor in CR0 configuration. Results of the MSM experiments are shown in Table 2. These results will be considered as reference reactivity values and will be used as a benchmark for the Area method.

**Table 2.** Reactivity of the subcritical configurations determined by the MSM method [6].

| Configuration | SC1/CR=0mm | SC1 | SC1/CR=600mm |
|---|---|---|---|
| Height of Control Rod 1 (mm) | 0 | 479.3 | 600 |
| Height of Control Rod 2 (mm) | 0 | 479.3 | 600 |
| MSM reactivity ($) | -6.35 ± 0.27 | -5.30 ± 0.23 | -5.09 ± 0.22 |

**The Area Method**

*Principle of the Area method*

When dealing with Pulsed Neutron Source (PNS) experiments, the Area method (also referred as the Sjöstrand method) [5] allows one to determine in a straightforward way the reactivity (in dollar) of a subcritical nuclear reactor with no input from theoretical calculations, as long as the assumptions of the neutron point kinetics hold in the reactor. This technique is based on the analysis of the time response of detectors placed in the reactor after a source neutron pulse. The evolution of the detector count rates strongly reflects that of the neutron population over time. Indeed, assuming that neutron point kinetics can represent the neutron population evolution over time, the equation of its time decrease after a pulse (considered as a Dirac peak) within the one-delayed neutron group approximation reads:

$$N(t) = N_0 \left[ \exp\left(\frac{\rho - \beta_{eff}}{\Lambda_{eff}} t\right) + \frac{\bar{\lambda} \Lambda_{eff} \beta_{eff}}{(\rho - \beta_{eff})^2} \exp\left(-\frac{\bar{\lambda} \rho}{\rho - \beta_{eff}} t\right) \right] \quad (1)$$

where $\bar{\lambda}$ is the average decay constant obtained by averaging the inverse constants $\frac{1}{\lambda_i}$. In equation (1), we can distinguish a "fast" component due to prompt neutrons, and a "slow" component, due to delayed neutrons. The integration of the prompt component over time gives the prompt surface $A_p$:

$$A_p = N_0 \frac{\Lambda_{eff}}{-\rho + \beta_{eff}} \quad (2)$$

whereas the integration of the delayed component gives the delayed surface $A_d$:

$$A_d = N_0 \frac{\beta_{eff} \Lambda_{eff}}{\rho(\rho - \beta_{eff})} \quad (3)$$

Then, the ratio of these two surfaces gives directly the value of the antireactivity in dollars:

$$-\rho_\$ = \frac{A_p}{A_d} = -\frac{\rho}{\beta_{eff}} \quad (4)$$

Experimentally, for a set of pulses repeated with a fixed frequency, a single Pulsed Neutron Source (PNS) histogram is constructed by summing the fission chamber time responses as a function of the time elapsed after the neutron pulse. The analysis consists in separating in this histogram the prompt neutron contribution from the delayed neutron one. After integrating the time spectrum to get the surfaces $A_p$ and $A_d$, the antireactivity can be calculated using Eq (4)

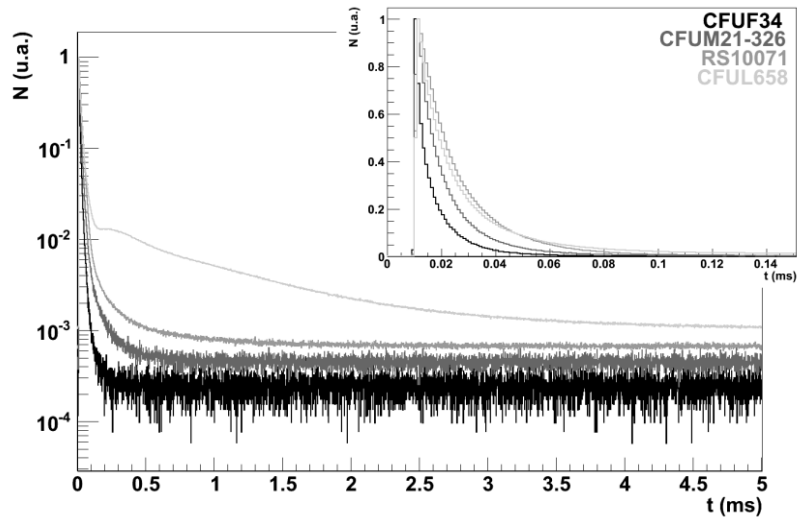

**Figure 3.** Time-dependent PNS histograms obtained with 4 different FCs for the reactor configuration SC1/CR=479.3 mm.

*Typical PNS histograms*

In order to extract the reactivity value of the SC1 and the SC1 variant configurations, the Area method was applied to the count rates measured during the PNS experiments by the ten FCs installed in the reactor. When necessary, count rates were corrected for dead time. Typical PNS histograms are presented in Figure 3 for various detector positions: CFUF34 in the core, CFUM21-326 at the core-reflector interface, RS100-71 in the corner of the 12x12 grid and CFUL658 inside the outer part of the reflector. These histograms were built by adding-up at least one million pulses for a beam frequency of 200 Hz and they are normalized to the same maximum.

Except for the CFUL658, we observe that the PNS time spectra have almost the same shapes, which however depend on the detector position inside the reactor (they are not homothetic). First, right after the neutron pulse injection, a sharp increase of the fission rates is observed. This delay, before reaching the maximum count rate, is explained by the neutron transport time from the source location all the way to the FC position. Then, the count rates decrease more or less rapidly, depending on the reactor region, within about 1.5 ms. This "fast component" corresponds to the prompt neutron driven decay of the neutron population. It is also observed that the closer to the reflector the FC, the slower the decay of this fast component. This behaviour might sign the presence of spatial effects, which are not predicted by the point kinetics model. Except for the two detectors located inside the outer lead reflector, beyond 2 ms, a quasi-constant level referred hereafter to as the delayed neutron level $L_d$, is reached. This so-called "slow component" is the sum of the contributions of the delayed neutrons originating from the successive pulses.

Obviously one must check that the neutron precursors have reached equilibrium before analysing the data within the framework of the Area Method. A study of the delayed neutron level saturation using point kinetics shows that at least 200000 pulses should be considered. Also, the PNS experiments should not be performed at frequencies larger than about f ≈ 500 Hz. Indeed, above this value, shorter time intervals between pulses would prevent the PNS histogram from reaching the delayed neutron level. Looking at Fig. 3, one can see, unfortunately, that this frequency upper limit becomes lower when a detector farther away from the core is considered.

The constant level of the delayed neutrons $L_d$ is first obtained by calculating the average count rate on a domain ranging from a fixed upper time limit, $t_{max}$, to a lower time limit, $t_{min}$. $t_{max}$ is simply the period between two beam pulses. The lower limit $t_{min}$ is chosen in the flat region of the PNS histogram in order to get a good estimate of $L_d$ even for the smaller FC (CFUF34) and in order to maintain the systematic error on $L_d$ around 1% for the FCs having the slowest prompt neutron population decrease. Finally $t_{min}$ was fixed to $t_{min} = t_{max} - 0.5$ ms. Then:

$$A_d = \int_{t=0}^{t_{max}} L_d \, dt = \frac{L_d}{f} \quad (5)$$

Introducing $A_{tot}$, the total number of counts in the PNS histogram, we have:

$$\frac{\rho}{\beta_{eff}} = -\frac{A_p}{A_d} = -\frac{A_{tot} - A_d}{A_d} = 1 - \frac{A_{tot}}{A_d} \quad (6)$$

This relationship is valid only if the neutron intrinsic source originating from the fuel can be neglected. It is the case here, since the fuel is metallic uranium and was never irradiated at high power.

**Results**

The Area method was applied to reaction rates measured by the ten fission chambers during the PNS experiments for the three different subcritical configurations obtained by moving the control rods. Figures 4 and 5 show the results. Reactivity values extracted according to formula (6) are represented by solid dots. The error bars were calculated by taking into account the statistical as well as systematic errors. The horizontal dashed line represents the reactivity of the subcritical configuration as inferred from the MSM method, while the solid horizontal lines show the uncertainty range on the MSM value.

One notices a dispersion of the results, which seem to depend on the detector location in the reactor. Three groups can be identified. The first one contains only the CFUF34 detector, which is the only one located in the reactor core. It is also the only one from which the reactivity value obtained with the Area method is in very good agreement with that of the MSM method. The second group gathers six (RS10074, RS100-71, CFUL659, CFUM326, CFUM325 and RS10072) or even seven (RS10075) detectors, which are located either at the core-reflector interface or in the corners of the 12x12 grid, in the inner part of the reflector. The last detectors (RS10075, CFUL653 and CFUL659) form the third group. They are located rather far away from the core, in the outer part of the reflector, outside the casing. Clearly the Area Method fails at providing the correct value of the reactivity when the FC are not in the core. The effect seems to be stronger when the detector is farther from the core. In the case of the third group, one just needs to look at Figure 3 to observe that the neutron population does not decay as predicted by neutron point kinetics. Furthermore the neutron population does not even reach the delayed neutron level within the time window corresponding to the period between the beam pulses. In these conditions, the area $A_d$ is overestimated, which leads to an underestimation of $A_p$ and the reactivity value extracted is wrong.

In order to correct for this detector location effect, we now turn to Monte Carlo simulations with MCNP [8]. Indeed, if the dispersion of the reactivity values given by the Area method is due to spatial effects, it should be possible to use Monte Carlo simulations of neutron pulses to correct for them since Monte Carlo simulations transport neutrons without geometric approximations. First an MCNP input file with a simplified geometry of the VENUS-F reactor was created in order to save computing time and investigate our hypothesis that the spatial corrections are not very sensitive to the details of the geometry. Second, Monte Carlo correction factors to be applied to the experimental values of reactivity can be calculated for each configuration and each detector location by:

$$f_{area} = \frac{\left(\frac{\rho^c}{\beta_{eff}^c}\right)}{\left(\frac{R_p^c}{R_d^c}\right)} = \left(\frac{\rho^c}{\beta_{eff}^c}\right)\left(\frac{R^c - R_p^c}{R_p^c}\right) \quad (7)$$

where $\rho^c$ is the reactivity computed with the MCNP model for the considered core configuration of VENUS-F. $R_d^c$ and $R_p^c$ are fission rates at some detector location, due to a Dirac pulse at the core center, associated with delayed neutrons and with prompt neutrons, respectively. Since MCNP cannot calculate the former, the total fission rate $R^c$ is computed and the difference $R^c - R_p^c$ is used instead. $\beta_{eff}^c$ is the calculated effective delayed neutron fraction associated with the reactor configuration. Since Monte Carlo estimates of $\beta_{eff}^c$ are very time consuming, this parameter was taken from calculations performed with the deterministic code ERANOS[9] for the same reactor configuration, which gave $\beta_{eff}^c$ =722 pcm [10]. $\rho^c/\beta_{eff}^c$ can be regarded as the "true" reactivity value, while $R_p^c/R_d^c$ is the distorted one corresponding to some detector position. If point kinetics would hold everywhere in the VENUS-F reactor, $f_{area}$ would be equal to one. Finally the corrected reactivity value reads:

$$\rho_\$ = f_{area} \times \frac{A_p}{A_d} \quad (8)$$

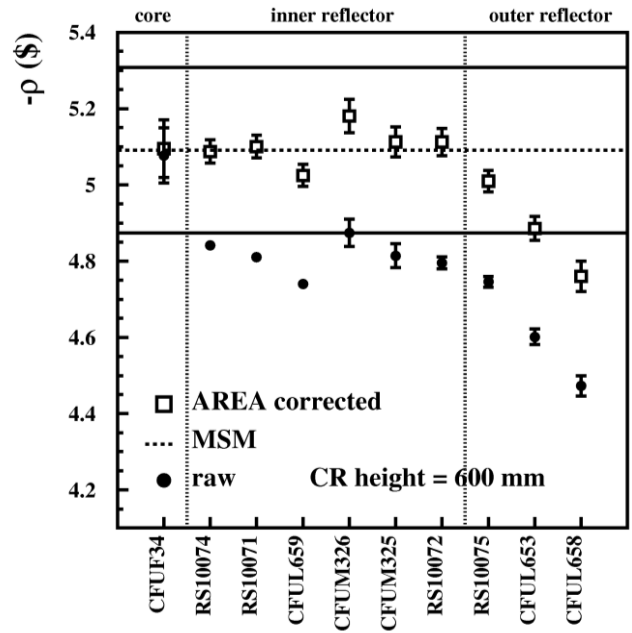

**Figure 4.** Uncorrected (solid dots) and corrected (open squares) reactivity values extracted from detector counts for the reactor configuration SC1/CR=600 mm. The MSM reference value is the dashed line and its uncertainty range is given by the solid lines.

The corrected values are symbolized by open squares on Figures 4 and 5. For every configuration, as expected, the effect of the correction is negligible for the CFUF34 located inside the core. Except for the fission chambers installed in the outer lead reflector, the corrected values are all compatible with the reactivity given by the MSM method. It is not surprising that the correction fails for the FCs in the outer reflector, since for these, the delayed neutron level could not be reached in the PNS time window given by the 200 Hz frequency of the beam.

Finally, discarding the results obtained for the fission chambers located in the outer part of the reflector, the average corrected value of reactivity was calculated for the three configurations studied. To calculate the uncertainty, it was assumed conservatively that the correlations are at maximum between the values given by the detectors. As can be seen in Table 3, the agreement between the MSM reactivity and that given by the Area Method is remarkable.

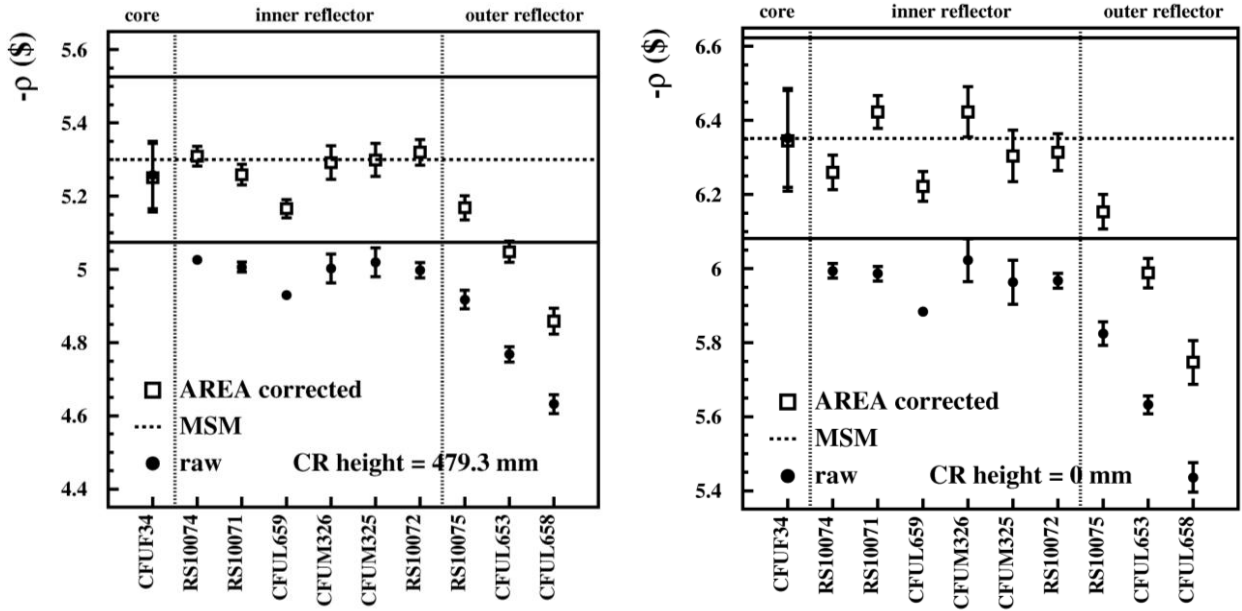

**Figure 5.** Same as Figure 4 but CR height at 479.3 mm (left) and CR height at 0 mm (right).

**Table 3.** Average reactivity value given by the Area method compared with the MSM reference value, for the three reactor configurations studied.

| CR height (mm) | $\langle\rho\rangle_\$^{Area}$ | $\langle\rho\rangle_\$^{MSM}$ |
|---|---|---|
| 600 | -5.09 ± 0.03 | -5.09 ± 0.22 |
| 479.3 | -5.26 ± 0.03 | -5.30 ± 0.23 |
| 0 | -6.31 ± 0.05 | -6.35 ± 0.27 |

**Conclusions**

In this paper, the reactivity estimates of three different subcritical levels of the VENUS-F reactorextracted from PNS experiments with the Area method were presented. First, the technique was applied to count rates measured by ten fission chambers used during PNS experiments driven by the GENEPI-3C deuteron accelerator and performed for three different reactivity levels of the reactor. The dispersion observed among the reactivity estimations inferred from the responses of the detectors spread over the entire reactor volume pointed out that space-energy effects bias the results and that they must be accounted for. Then we exposed the methodused to compute, by means of simulations performed with MCNP, correction factors for all the detector positions inside the VENUS-F reactor. Except for two fission chambers located inside the outer lead reflector, all the corrected reactivity values were compatible and in good agreement with the reference values previously estimated with the MSM method.

**Acknowledgements**

This work is partially supported by the 6th and 7th Framework Programs of the European Commission (EURATOM) through the EUROTRANS-IP contract # FI6W-CT-2005-516520 and FREYA contract # 269665, and the French PACEN program of CNRS. The authors want to thank the VENUS reactor and GENEPI-3C accelerator technical teams for their help and support during experiments. They are also very grateful to the physics control service of SCK-CEN.